\documentclass[12pt,a4paper,final]{iopart}
\usepackage{iopams}

\usepackage{graphicx}
\usepackage[breaklinks=true,colorlinks=true,linkcolor=blue,urlcolor=blue,citecolor=blue]{hyperref}

\begin{document}

\title[Light field integration in SUGRA theories]{Light field integration in SUGRA theories}

\author{Diego Gallego}
\address{Escuela de F\'isica, Universidad Pedag\'ogica y Tecnol\'ogica de Colombia (UPTC),\\
Avenida Central del Norte, Tunja, Colombia.}

\ead{diego.gallego@uptc.edu.co}

%

\begin{abstract}
We revisit the integration of fields in ${\cal N}=1$ Supergravity
with the requirement that the effective theory has a reliable
two-derivative supersymmetric description. In particular we study,
in a supersymmetric manifest way, the situation where the fields
that are mapped out have masses comparable to the Supersymmetry
breaking scale and masses of the remaining fields.
\\
We find that as long as one stands in regions of the field
configuration space where the analytic continuation to superspace of
the F-flatness conditions be reliable equations of motion for the
fields that are being mapped out, and provided their solutions are
stable regardless the dynamics of the remaining fields, such a
two-derivative description is a reliable truncation of the full
effective theory.
\\
The study is mainly focused to models with two chiral sectors, $H$
and $L$, described by a K\"ahler invariant function with schematic
dependencies of the form $G=G_H(H,\bar H)+G_L(L,\bar L)$, which
leads to a nearly decoupled theory that allows the previous
requirements to be easily satisfied in a consistent way.
Interestingly enough for the matters of our study this kind of
models present an scenario that is as safe as the one presented in
sequestered models.
\\
It is also possible to allow gauge symmetries as long as these
appear also factorized in hidden and visible sectors. Then, the
integration of the hidden vector superfields is compulsory and
proceeds reliably through the D-flatness condition analytically
continued to superspace.
\end{abstract}

\vspace{2pc}

\section{\label{intro}Introduction}

Supersymmetry (SUSY) and in general Supergravity (SUGRA) not only
continues to be the preferred playground for models beyond the
Standard Model, but also an ideal framework for dealing with
situations where otherwise many calculations would be either
impossible or unreliable. However, any constructed model has in mind
only a small subset of the entire bunch of fields present in
explicit realizations, and these are regarded as encoding all the
important dynamics under study. Physically what one has in mind is
that the rest of the fields are either decoupled or that their
dynamics are negligible. Formally the neglected fields are supposed
to be integrated out in such a way that the resulting theory is, at
least approximately, SUSY.
\\
Integrating out fields in ${\cal N}=1$ SUGRA theories led recently
to some discussion
\cite{deAlwis:2005tg,deAlwis:2005tf,Abe:2006xi,Achucarro:2008sy,Achucarro:2008fk,Choi:2008hn,Choi:2009jn,conMarcoI,Lawrence:2008ar,conMarcoII}
settled finally by the work of Brizi, Gomez-Reino and Scrucca
\cite{Brizi:2009nn} where, by requiring an effective two-derivative
SUSY description, approximate superfield equations of motion
(e.o.m.) were derived, for the fields to be integrated out, together
with the estimate of the deviations from the exact effective higher
order theory. These can be understood in the light of a low-energy
effective theory where higher order terms appear suppressed by the
mass of the fields being integrated out and, therefore, turn out to
be subleading. A general result of the work by Brizi et al. is that
the gravitational effects to the e.o.m. are automatically negligible
once the masses of the integrated fields lie far above the
characteristic energies of the effective theory, which include now
the SUSY breaking scale, and therefore the leading superfield e.o.m.
coincide with the ones of rigid SUSY.
\\
There are, however, scenarios where one might like to get rid of
some fields despite the fact that no hierarchy is realized. Already
in ordinary field theories it is clear that in such a case higher
order derivative terms are no longer suppressed, as the kinetic
energies in the effective theory are comparable to the masses of the
integrated fields. An obvious situation that circumvents this
problem is the case where both sectors, the one to be integrated out
and the one to be kept, denoted hereafter by $\{H\}$ and $\{L\}$
respectively, are completely decoupled. For rigid ${\cal N}=1$ SUSY,
without vectors fields, this is obtained for K\"ahler potential and
superpotential factorized schematically as follows:
\begin{equation}\label{factoRigidSUSY}
K=K_H(H)+K_L(L)\,,~~W=W_H(H)+W_L(L)\,.
\end{equation}
In SUGRA, instead, the theory is described by the generalized
K\"ahler invariant function, $G=K+\ln|W|^2$, so the previous
factorization does not hold in $G$ nor in the theory. Moreover,
gravitational interactions implies that even if $G$ turns out to
have a factorized form, i.e., $G=G_H(H)+G_L(L)$, the Lagrangian has
not a fully decoupled structure as can be seen already in the scalar
potential,
\begin{equation}
V=e^G\left(G^{I\bar J}G_IG_{\bar J}-3 \right)\,,
\end{equation}
with $G^{I\bar J}\equiv \big(G_{I\bar J})^{-1}$ the inverse scalar
manifold metric, the subindex $I$ denoting derivatives respect to
the superfield $\phi^I$, and everything is evaluated in the lowest
component of the superfields. In fact, in this context the most
ideal scenario would be the sequestered  models
\cite{Randall:1998uk, Dimopoulos:2001ui,Cheung:2010mc} where only
gravitational interactions enter in the interplay between the
sectors.
\\
A factorizable $G$ function leads still to some decoupling, as was
first discussed in \cite{Binetruy:2004hh} and later on studied in
\cite{Achucarro:2008sy,Achucarro:2008fk,conMarcoI}. A more detailed
study was done by the author in \cite{mioLVS}, where also vector
fields in simple setups where included, but still following a
component approach where SUSY is not manifest in the prescription
and, therefore, it is not completely clear how to understand the
results in a fully SUSY framework. Moreover, it is not clear the
extend of validity of the constrains found there as the e.o.m. are
not actually fully scanned.
\\
The present letter follows closely the analysis in
\cite{Brizi:2009nn}, now for nearly decoupled theories and no
hierarchies in the game, looking for the conditions for a reliable
two-derivative SUSY description in the effective theory. We find
that such a description exist and is reliable as long as the
analytic continuation of the F-flatness condition to the superspace
is a reliable e.o.m. for the fields that are being integrated out
and has stable solutions, regardless the dynamics of the field
sector kept in the theory.
\\
The superfield approach allows a neat analysis in the case of
presence of gauge symmetries. In this case, in order the $H$ sector
not to be sourced back, the fields to be integrated out can be
charged only under some hidden gauge group whose gauge kinetic
function dependency on the $L$ sector should be suppressed. In the
same way the gauge kinetic function of the visible sector depends on
the $H$ fields in a mild way. Then, the leading superfield e.o.m.'s
are the analytic continuation of the F-flatness conditions plus the
D-flatness ones, leading to a two-derivative SUSY theory.
\\
Our results support and generalize the findings of \cite{mioLVS}.
Working directly in the superspace, however, allows us to spot
directly the fact that it is not necessary to restrict artificially
to slow varying solutions being in fact a requirement coming from
the e.o.m., as can be also understood from SUSY transformations.
These findings are particularly relevant in the context of SUSY
breaking scenarios, and the related issue of moduli stabilization in
Superstring/M-theory, where most of the fields are regarded as SUSY
preserving and a detailed description of the SUSY breaking and
moduli stabilization is performed only on a tiny subset of fields.
In particular the seminal work of Kachru, Kallosh, Linde and Trivedi
\cite{KKLT} falls in the kind of scenarios where a hierarchy,
dictated by the ratio between the flux and non-perturbative
dynamics, is present and therefore the results of Brizi et al.
apply. On the other hand, for natural Vacuum Expectation Value (VEV)
of the superpotential, i.e., $\langle W\rangle\sim 1$ in Planck
units, all fields get important, and of the same order, gravity
contributions to the masses and therefore no hierarchy is realized.
Low energy SUSY is still possible if the VEV for the $G$ function is
negatively large thanks to the universal factor $e^G$ in the
potential. This is what precisely happens in the so called Large
volume scenarios (LVS) for type-IIB Superstring compactifications
\cite{Balasubramanian:2005zx,Cicoli:2008va} where, moreover, the
coupling between the Kh\"ahler moduli, $T$, and the SUSY preserving
dilaton and complex structure moduli, denoted by $U$, is described
by
\begin{equation}
G_{mix}\sim\frac{\xi(U,\bar U)}{{\cal V}(T,\overline
T)}+\frac{W_{np}(T,U)}{W_{flux}(U)}+h.c.\,,
\end{equation}
with $\xi$ a function of the dilaton resulting from $\alpha^\prime$
corrections \cite{Becker:2002nn} and $W_{flux}\gg W_{np}$ the flux
induced and non-perturbative parts in the superpotential. Thus, for
large values of the compact manifold volume, ${\cal V}$, the $G$
function realizes an approximate factorizable form and our results
apply (for details and numerical examples see \cite{mioLVS}).
\\
We should mention that, although with some broad applicability in
moduli stabilization models, our analysis should be repeated for
scenarios where higher order operators are relevant and the
two-derivative level leads to poor descriptions, like the case of
cosmological models of inflation where the background dynamics
should be taken into account \cite{EffInfl,Tolley:2009fg}. Then, it
is necessary to keep full track of the higher order operators to get
insights of the effective SUGRA theory \cite{Baumann:2011nm}.
Nevertheless, some analyses are valid at this level
\cite{Postma,Achucarro:facto,Achucarro:2010jv,Hardeman,Achucarro:2011yc,Cespedes:2012hu,Achucarro:2012sm}.
\\
The letter is organized as follows: section two is dedicated to
review the general procedure of integrating out fields defining what
we call the effective description, where the degrees of freedom that
are mapped out are not heavier than the ones kept in the theory, in
contrast with the usual low energy effective theories. In section
three the arguments in \cite{Brizi:2009nn} are reviewed introducing
SUSY as a global symmetry. Here a first instance of models with a
reliable two-derivative SUSY effective description is shown. In
section four gravity enters in the game regarding only chiral
superfields. In here, after recovering results in
\cite{Brizi:2009nn}, we show how the factorizable models can have
such an effective description and the superfield e.o.m. to use in
this case. Section five is dedicated to study the case where gauge
symmetries are present, exploring also the possibility of having a
charged hidden sector. The last section discusses the gravitational
terms and the gauge fixing of the superconformal symmetry, an issue
not regarded in previous studies. We close with some summary and
discussion of the results.

\section{Integrating out fields and effective descriptions}

Let us consider a field theory with two kind of modes, $H$ and $L$,
described by an action $S[H,L]$. Suppose we are in a situation where
only the $L$ modes can be realized in the initial and final states.
Then, the dynamics of the $L$ fields can be described by a theory
that does not depend on the $H$ ones. This theory is the result of
summing up over the $H$ intermediate states, in a procedure that in
the path integral formalism goes precisely as integration over the
$H$ modes, defining the effective action, $S_{Eff}$, for the
remaining fields,
\begin{equation}
e^{-S_{Eff}[L]}=\int [dH]e^{-S[H,L]}\,,
\end{equation}
where we use the Euclidean space notation. For practical purposes we
expand around the classical solution for the $H$ fields so the path
integral is now over quantum fluctuations,
\begin{equation}
e^{-S_{Eff}[L]}=e^{-S[H_0,L]} \int [d \delta
H]\exp\left\{-\frac{\delta^2 S[H,L]}{\delta H\delta H}\delta H\delta
H+\cdots\right\}\Big|_{H=H_0}\,,
\end{equation}
with $H_{0}$ the solutions to the classical e.o.m. $\frac{\delta
S}{\delta H}=0$, and the ellipses containing higher order terms in
the quantum fluctuations. Notice, however, that the solutions to
this classical e.o.m. are expected to depend on the $L$ fields for
which quantum fluctuations are still on. Therefore, in general $H_0$
is not the classical solution for $H$.\footnote{The classical e.o.m.
is in fact more general once the action that is originally taken
contains quantum
corrections like non-perturbative effects. 
} At this level, then, the effective action is given by
\begin{equation}
S_{Eff}[L]=S[H_0,L]\,.
\end{equation}
So far we have not specified what we mean by the two kinds of modes.
In the usual low energy effective theories the $L$ are low energy
modes, and the $H$ high energy ones, where the distinction is made
by some energy scale $\Lambda$. In this case then the path integral
defining the effective action is restricted to modes with momentum
higher to $\Lambda$. Formally, however, the procedure can be applied
to get rid of any kind of mode, as long as one ensures that in the
asymptotic states only $L$ fields appear. For example if the action
turns out to be the sum of actions for each sector, i.e.,  the two
sector are decoupled, and in the initial states only one kind of
fields appear. The situation we will deal with is of this kind,
where despite the fact the decoupling is not complete any mixing
term in the action will be parametrically small. This is what we
will call an effective description to be distinguished from the low
energy effective one.
\\
In general the theory that one obtains by the procedure above is a
higher derivative theory, and so suffers the pathologies of such
\cite{Pais:1950za}. Thus one might like to keep, consistently, only
up to two derivative operators though higher order operator in the
fields can be admitted. This reduces to consider in the e.o.m. only
the contribution from the potential disregarding the kinetic terms.
The truncation can be stated more precisely using the general form
the effective Lagrangian of the resulting theory \cite{Pich:1998xt},
\begin{equation}\label{genEFF}
 {\cal L}=\sum_i\frac{c_i}{\Lambda^{d_i-4}}{\cal O}_i\,,
\end{equation}
where $c_i$ are dimensionless couplings known as Wilson
coefficients, $\Lambda$ is the cut-off scale and ${\cal O}_i$ are
operators with dimension $d_i$. We see, therefore, that for low
energy effective theories higher order operators are naturally
suppressed by powers of the energy scale $\Lambda$ and the
requirement above is automatic.\footnote{Part of this analysis
clearly relies on the assumption that the Wilson coefficients have
no anomalous small or large values.} For the effective descriptions,
instead, the consistency of the truncation procedure relies on the
smallness of the couplings between the two sectors, which translates
in small Wilson coefficients for operators not present in the
original theory.
\\
Another consideration behind the philosophy of effective theories is
to ensure that the fields that have been mapped out be not sourced
back by any process involving the $L$ fields. For the low energy
effective theories this is guaranteed kinematically since the
effective theory cannot be used for describing processes with
energies comparable or larger than the masses of the $H$ fields. In
the case of effective description the constraint is dynamical,
coming from the decoupling. So the $L$ fields cannot excite $H$
modes.

\section{Global SUSY effective theories}\label{RigidSect}

The introduction of SUSY, and the requirement of a SUSY effective
description introduces a further issue, as was noticed in
\cite{Brizi:2009nn}: the usual two-derivative truncation for an
effective description of a field theory is not enough when SUSY is
implied, as higher order terms in the spinor bilinears and auxiliary
fields are mapped, by SUSY transformations, to higher order
derivative terms. Therefore, a further truncation in spinor
bilinears and auxiliary fields should be imposed which will be
reliable only if the missing terms are negligible. At the superfield
level this means neglecting SUSY covariant derivatives in the
K\"ahler potential and superpotential in the effective description,
as should be clear from the fact that these derivatives have as
components, both, space-time derivatives as well the spinor and
auxiliary components of the fields, these last ones encoding the
SUSY breaking energy scale. In other words, the solutions to the
superfield e.o.m., for the fields that are being mapped out, should
correspond either to field configurations where all the SUSY
covariant derivatives are negligible, or such that are independent
of any non-negligible one.
\\
Let us explore better the situation for a global SUSY theory and
consider models with two sectors of chiral fields $\{H^i\}$ and
$\{L^\alpha\}$ (notice the distinction in the indices). The exact
e.o.m. for the superfields $H$ are obtained from the generic
two-derivative Lagrangian
\begin{equation}
{\cal L}=\int d^2\theta d^2\bar \theta K+\int d^2\theta W+h.c.\,.
\end{equation}
The application of the variational principle in this action should
take into account the constraint $\overline {\cal D} \Psi=0$, with
${\cal D}_\alpha=\frac{\partial}{\partial\theta^\alpha}+i
\sigma^\mu_{\alpha\dot\beta}\bar \theta^{\dot\beta}\partial_\mu$ the
supercovariant derivative, on the chiral
superfields.\footnote{Following conventions and notations like in
\cite{Bilal:2001nv}.} To make this explicit we write the D-term part
of the action as a F-term one using the identity
\begin{equation}\label{DtoF}
\int d^2 \theta d^2\bar \theta G(\Psi,\bar \Psi)=-\frac14\int d^2
\theta  \overline {\cal D}^2 G(\Psi,\bar \Psi)\,,
\end{equation}
where ${\cal D}^2 =\overline {\cal D}_{\dot \alpha}\overline {\cal
D}^{\dot \alpha}$, the two expressions differing by total
derivatives. With this consideration the superfield e.o.m. for the
$H^i$ reads:\footnote{Along the paper we use the Latin subindex
notation to denote derivatives with respect superfields, e.g.,
$W_i\equiv\frac{\partial W}{\partial H^i}$.}
\begin{equation}\label{SUSYExact}
W_i-\frac14\overline{\cal D}^2 K_i=0\,.
\end{equation}
It is usually considered (see \cite{Affleck:1983mk}, also
\cite{Brizi:2009nn,Pomarol:1995np}) that, around the solutions to
this e.o.m., the energy scale associated to the second non-mixed
holomorphic derivatives of the superpotential, i.e., $W_{ij}$,
dominates over all others, say the ones associated to the
superpotential itself, pure $L$-sector and mixed derivatives, e.g.,
$W_{\alpha\beta}$ and $W_{i\alpha}$,  the space-time derivatives on
the fields and the auxiliary fields VEV's. With a regular behavior
in the K\"ahler potential this implies that the leading term in
(\ref{SUSYExact}) is the first one so the approximate e.o.m. reads:
\begin{equation}\label{heavycase}
W_i=0\,,
\end{equation}
which leads to a two-derivative SUSY description for the $L$ fields
as no SUSY covariant derivative is present. In particular the
solutions to this e.o.m. are vanishing $H$ auxiliary fields implying
no contribution to the SUSY breaking from the $H$ sector at leading
order.
\\
Physically the fact that the holomorphic $W_{ij}$ derivatives
dominate means that the masses of the $H$ fields, $M_H\sim W_{ij}$,
are larger than the remaining energy scales, namely, the masses and
kinetic energy of the $L$-sector fields as well the SUSY breaking
scale. Then, the theory obtained from this leading e.o.m. coincides
with the full higher order operator effective theory at first order
in an expansion in derivatives, spinor bilinears and auxiliary
fields, with the missing terms suppressed by $M_H$, precisely like
in any standard low-energy effective description.
\\
A second possibility is one where the dynamics ruling both sectors
are of the same order and therefore no significant hierarchies
appear. Still one might like to get rid of one sector, in which case
one should consider the situation of factorization of the action,
where the K\"ahler potential and superpotential have the form in
(\ref{factoRigidSUSY}), so the e.o.m. reads
\begin{equation}\label{leadingEOMFACTRIGID}
W_{H,i}(H)-\frac14\overline{\cal D}^2 K_{H,i}(H,\bar H)=0\,.
\end{equation}
Then, the integration of the $H$ sector is completely arbitrary as
there is not $L$ field dependence on the e.o.m, so the solutions
would not affect the $L$ sector dynamics, which in particular
continue to be described by a two-derivative SUSY theory. Still, let
us explore the possibility of neglecting the second part of the
equation, such that the e.o.m. would be $W_{H,i}=0$. The components
of this equation are given in \ref{EOMComps} where is also shown
that there is a trivial solution, i.e., vanishing spinor and
auxiliary components and null space-time derivatives. Nicely enough
these are precisely the components of the covariant derivatives
appearing in the second part of the e.o.m. and, therefore, this kind
of solutions are also solutions of (\ref{leadingEOMFACTRIGID}).
Thus, the integration is well described by an e.o.m where the
supercovariant derivatives are simply not considered.
\\
This kind of solutions to the e.o.m are SUSY preserving, a
requirement that in presence of gravity, as we will see, is a
compulsory one. Another motivation to stick to the SUSY preserving
solution is that we can generalize our analysis by including a small
mixing, i.e., $W_{mix}(H,L)$ or $K_{mix}(L,\overline L,H,\overline
H)$, that makes the decoupling not exact. In this case, if we
parametrize the magnitude of the mixing terms by $\epsilon\ll1$ we
expect deviations to the previous solutions of this order. Then, the
approximated e.o.m. is reliable up to terms of ${\cal O}(\epsilon)$
and the exact solutions are given by $H=H_0+\delta H$, with $H_0$
the solution to $W_{H,i}=0$ and
\begin{equation}
\fl\delta H^{i}=\epsilon\, O^{ij}\left[\frac14 \overline{\cal D}^2
K_{mix,j}- W_{mix,j}+\frac14 K_{H,j\bar j}\overline{ W}_H^{\bar
j\bar i}\overline{\cal D}^2\left(\frac14 {\cal D}^2K_{mix,\bar
i}-\overline{W}_{mix,\bar i}\right)\right]\Big|_{H_0}+{\cal
O}(\epsilon^2)\,,
\end{equation}
with $O^{ij}\equiv (W_{H,ij}-\frac{1}{16} K_{H,i\bar j}\overline{
W}^{\bar j\bar i}_HK_{H,\bar ij}\overline{\cal D}^2 {\cal
D}^2)^{-1}$ and $\overline{ W}^{\bar j\bar i}_H\equiv
\left(\overline{ W}_{H,\bar j\bar i}\right)^{-1}$. Notice that the
mixing term induces dependency of the solution on SUSY derivatives
on the $L$ superfields, which by no means we can suppress {\sl a
priori}. It is thanks to the small parameter that these are
controlled and the whole $\delta H$ is small. Plugging the solution
back in the original theory induces, schematically,
\begin{eqnarray}
W_{eff}&=&W(H_0)+ \frac12 \epsilon^2 W_{ij}(H_0)\delta H^i \delta H^j+{\cal O}(\epsilon^3)\,,\\
K_{eff}&=&K(H_0)+ \epsilon K_i(H_0)\delta H^i+ \epsilon K_{\bar
i}(H_0)\delta \bar H^{\bar i}+{\cal O}(\epsilon^2)\,,
\end{eqnarray}
where we have factored out the parameter $\epsilon$ characterizing
the magnitude of $\delta H$. The ${\cal O}(\epsilon)$ corrections to
the K\"ahler potential at first sight seem to affect drastically the
theory by inducing terms with size comparable to that of the
original ones, in particular the terms in $K_{mix}$. However, since
$H_0$ is a constant c-number and $\delta H$ is a chiral superfield,
the corrections correspond to a K\"ahler transformation and leads
only to total derivatives in the Lagrangian. We conclude, therefore,
that since the supercovariant derivatives appear from ${\cal
O}(\epsilon^2)$, a truncation at the ${\cal O}(\epsilon)$ level
leaves an effective description that has the structure of a
two-derivative SUSY theory.
\\
We have, then, that in both cases the superfield equation to be used
in the integration of fields is the promotion of the F-flatness
condition to the superspace level. We can state this in the
effective theory language, where always is only a small portion of
the field configuration space that is explored, by saying that as
long we stand in a region where the field configuration for the $H$
sector is approximately SUSY preserving, say with deviations
parametrically of order $\epsilon$, the theory obtained by
integrating out this sector is a theory that at leading order in
$\epsilon$ is a two-derivative SUSY theory.\footnote{The small
parameter for the case with hierarchies is given by the ratio of
mass scales, i.e., $\epsilon\sim m_L/m_H$.} For the case of
factorizable models the reliability goes up to the ${\cal
O}(\epsilon)$.
\\
One should expect that once the full decoupling is loosed further
constrains might proceed. This since even in the case of small
coupling after long periods of time, and space variations, the two
sectors share enough energy to affect one to the other.\footnote{We
thank an anonymous referee for pointing out this important point.}
This constrain is however already encoded in the fact that the
solution for the $H$ sector has leading part with vanishing
space-time derivatives, implying that for the $L$ sector the $H$ one
is an homogeneous one from which such an energy transfer does not
occur. But once the coupling is allowed to be of order $\epsilon$,
the space-time variations are also of this order and therefore the
reliability of the two-derivative description holds for space-time
scales of order $\epsilon^{-1}$. In this case of SUSY theories, this
will be also the space-time scale for which a SUSY solution is
reliable, for the $H$ auxiliary components start to be non
negligible. The same considerations apply for the following SUGRA
case.
\\
A final remark is in order. Although a hierarchy or decoupling is a
necessary condition for a small perturbation on the solution for the
$H$ sector, which we need to preserve SUSY, a further assumption is
required, namely that the $H$ sector be indeed stabilized.
Understanding by stabilization points the ones where the Hessian of
the scalar potential has no negative nor zero eigenvalues in these
directions, so that indeed these acquire positive mass squared and
so fluctuations are not dramatic for the stability of the solution.

\section{Two-derivative SUGRA effective theories}

We work directly with the K\"ahler invariant function, $G=
K+\ln|W|^2$, as this K\"ahler gauge is usually cleaner in the
results and therefore convenient for cases where the superpotential
is non vanishing or, as in our case, does not introduce any
important scaling by, say, a tiny VEV. It is also convenient to use
the superconformal formalism and compensator technique to write down
the action \cite{Kaku:1978nz,VanNieuwenhuizen:1981ae,Kugo:1982cu}.
In this setup the off-shell minimal SUGRA supermultiplet is split
and one of the two auxiliary fields is now contained in a
compensator chiral supermultiplet $\Phi$, required by Weyl symmetry,
which later on is gauge fixed in order to recover the actual
symmetries of SUGRA. Under this formalism the tensor calculi are
almost the same of rigid SUSY, allowing to write down the Lagrangian
as an integral over rigid supercoordinates. In our K\"ahler gauge,
for the moment without gauge interactions, the generic
two-derivative Lagrangian reads \cite{Kugo:1982cu,Ferrara:1983dh}:
\begin{equation}\label{SUGRAL}
{\cal L}=-3\int d^2\theta d^2\bar\theta e^{-G/3}\Phi \bar \Phi+\int
d^2\theta\Phi^3+h.c.+\cdots\,,
\end{equation}
the ellipses containing terms implying the graviton, gravitino and
the remaining auxiliary field from the SUGRA multiplet, also
including couplings with the matter multiplets. For the moment we
neglect them in our analysis and comment about the consistency of
the procedure at the end.
\\
Again we consider models with two sectors of chiral fields $\{H^i\}$
and $\{L^\alpha\}$, then the exact superfield e.o.m. for the $H^i$
reads:
\begin{equation}\label{fullEOM1}
-\frac14\Phi\overline{\cal D}^2\left( G_i e^{-G/3}\bar
\Phi\right)=0\,,
\end{equation}
where we have used again the identity (\ref{DtoF}) and the fact that
the superfield $\Phi$ is chiral. Regarding $\Phi\neq 0$ and
expanding the previous expression we have,
\begin{equation}\label{fullEOM}
\fl e^{-G/3}\bar \Phi\left( G_{i\bar I\bar J}\overline {\cal
D}\bar\phi^{\bar I}\overline {\cal D}\bar\phi^{\bar J}+G_{i\bar
I}\overline {\cal D}^2\bar\phi^{\bar J} \right)+G_i\overline {\cal
D}^2\left( e^{-G/3}\bar \Phi\right)+2 G_{i\bar I}\overline{\cal
D}\bar\phi^{\bar I} \overline {\cal D}\left( e^{-G/3}\bar
\Phi\right)=0\,,
\end{equation}
where for simplicity in the notation we omit the spinor index in the
SUSY covariant derivatives, and the $I$, $J$ indices run over all
superfields $H^i$ and $L^\alpha$. From previous arguments, the SUSY
two-derivative description is reliable if somehow around the
solution to the e.o.m. we can neglect the covariant derivatives.
\\
Like before we can consider the case where the fields to be
integrated out are heavy compared with the other energy scales (see
for  example \cite{Choi:2008hn,Choi:2009jn,conMarcoI,Brizi:2009nn})
in which case the e.o.m. is dominated by the term proportional to
$G_i$, whose leading part is $W_i/W$. Then at leading order in
inverse powers of the heavy masses the solution satisfy the equation
$W_i=0$ which is precisely the one found for the case of rigid SUSY,
and not depending on the SUSY covariant derivative leads to a
reliable two-derivative SUGRA theory.
\\
One can as well study the case of no hierarchy in the masses with
decoupled sectors, though in SUGRA exact decoupling is not possible
due to the gravitational interactions. We leave the most ideal case
of sequestered models for a later comment, and concentrate in a
second class of models, proposed first in \cite{Binetruy:2004hh} and
studied at the level of the scalar potential in
\cite{Achucarro:2008sy, conMarcoI, mioLVS}. The main property of
such models can be summarized in a K\"ahler invariant function with
the following structure
\begin{equation}\label{factG}
G=G_H(H,\bar H)+G_L(L,\bar L)+\epsilon\, G_{mix}(H,\bar H,L,\bar
L)\,,
\end{equation}
with $G_H$ and $G_L$ of the same order of magnitude and $\epsilon$
small, parameterizing the coupling between the two sectors. We
emphasize, however, that the smallness of the mixing is not
necessarily due to a small coupling but rather that around the
solutions to the e.o.m. all mixed terms turn out to be small. This
form for the $G$ function in the superfield e.o.m. implies  that all
mixed derivatives, $G_{i\alpha}$ or higher order, are suppressed so
that the leading terms in (\ref{fullEOM}) are proportional either to
covariant derivatives of the $H$ fields or to the $G_i$.
Schematically this is:
\begin{equation}\label{leadEOMwithep}
G_i\overline {\cal D}^2\left( e^{-G/3}\bar \Phi\right)+{\cal
O}\Big(\overline {{\cal D} H},\big(\overline {{\cal D}
H}\big)^2,\overline{\cal D}^2\bar H\Big)={\cal O}(\epsilon)\,.
\end{equation}
For solutions with an approximate two-derivative SUSY description in
the effective theory the second term should be negligible and, thus,
the first one must vanish independently.\footnote{A SUSY description
is possible even in the case the hidden sector breaks SUSY as long a
superfield description is taken for the goldstino (see
\cite{Choi:2009jn}). The symmetry, however, will be non-linearly
realized.} In general the factor accompanying the $G_i$ is non zero
since depends on the supercovariant derivatives on the $L$ fields,
which are expected to be large and supposed to be linear
independent, then this term vanishes only if we require $G_i=0$. Due
to the factorization at leading order in $\epsilon$ the $G_i$
composite superfield depends only on the $H$ components (see
\ref{EOMComps}) and it vanishes trivially if the spinor and
auxiliary components of the solution are null, as well the
derivatives for the lowest and spinor components. The lowest
component in turn should satisfy the F-flatness condition in the $H$
directions. In fact, the e.o.m. we are finding is noting but the
analytical extension of these F-flatness conditions to superspace.
Nicely enough, the supercovariant derivative acting on these
solution automatically vanish. We conclude, therefore, that at
leading order in $\epsilon$ the exact e.o.m. is solved by the
solutions to the equation
\begin{equation}\label{leadingSol}
G_i=0\,.
\end{equation}
In particular, for heavy fields, compared with the SUSY breaking
scale, this e.o.m. reduces to the one found before,
eq.(\ref{heavycase}).
\\
Being more explicit, the exact solution for the $H$ superfields has
the following schematic form:
\begin{equation}\label{exactsolution}
H=H_o+\epsilon \tilde H(L,\bar L,\bar \Phi,\overline{\cal D}\bar
L,\overline{\cal D}\bar \Phi)\,,
\end{equation}
where $H_o$ is the solution to $\partial_iG_H=0$ and the remaining
encodes the non-constant and non-holomorphic part, which in case of
not being suppressed would spoil the two-derivative SUSY
description. Plugging back the solution into the K\"ahler invariant
function, we have that the effective theory is described by
\begin{equation}\label{Geff}
G_{eff}=G_{H,o}+G_L(L,\bar L)+\epsilon G_{mix,o}(L,\bar L)+{\cal
O}(\epsilon^2)\,,
\end{equation}
with the ``nought'' label indicating evaluation at $H=H_o$. Here it
is clear that the theory is described, up to next to leading order
in $\epsilon$, by a valid $G$ function with no supercovariant
derivatives and therefore has a reliable two-derivative SUSY
effective description.
\\
We can apply the analysis to the sequestered case, for which the Lagrangian is given by \cite{Randall:1998uk, Dimopoulos:2001ui}
\begin{equation}\label{seq}
\fl {\cal L}=-3\int d^2\theta d^2\bar\theta e^{-G_H/3}\Phi \bar
\Phi-3\int d^2\theta d^2\bar\theta e^{-G_L/3}\Phi \bar \Phi+\int
d^2\theta\Phi^3+h.c.+\cdots\,,
\end{equation}
where $G_H$ and $G_L$ depend only on $H$ and $L$ respectively. Here
it is clear that due to gravitational interactions, encoded in the
compensator, the two sectors cannot be completely decoupled, though
the situation is better than in the previous case. The e.o.m. for
the $H$ superfield has an analogous expression like (\ref{fullEOM}),
replacing $G$ by $G_H$ and the index $I$, $J$ running only over the
$H$ sector. Then, the analysis follows almost verbatim noticing that
now the corrections to the exact SUSY condition depend only on the
compensator, which plays the role of communicating any SUSY breaking
effect from, or to, the visible sector in what is called gravitino
mediation part of the full anomaly mediation going on in this kind
of models \cite{Eramo:2012qd}. We find, therefore, that for the
effects of having a two-derivative SUSY effective description the
factorizable models are as safe as the sequestered ones, although
the last ones have the advantage of being further decoupled
strengthening the dynamical constraint on the excitation of $H$
fields.
\\
Let us close this section by drawing attention to a potential issue
on equation (\ref{leadingSol}), that is the fact that, contrary to
the exact e.o.m (\ref{fullEOM1}), it is not a chiral superfield
equation. Indeed the $G$ composite superfield is real and therefore
the equation has more components than a chiral one, preventing us,
in general, from using it for the integration of chiral fields
\cite{Brizi:2009nn}. In the case of factorizable models, however,
this is avoided as the antiholomorphic components of the equations
are trivially consistent with the holomorphic ones, in the sense
that both lead to vanishing spinor, spinor derivatives and auxiliary
components in both $H$ and $\bar H$, which at the same time are
consistent with the lowest component of the equation that, as said
before, is the F-flatness condition. Therefore, the leading part of
the solution is given by the chiral set $H_o=\{h_o,0,0\}$.

\section{Gauge interactions}

The presence of gauge interactions modifies the analysis, first by
the inclusion of the vector superfields $V^A$ in a gauge invariant
way in $G=G(\phi^I,\bar \phi^{\bar I}, V^A)$, the index $A$ running
over the gauge group generators, and then by their kinetic term,
\begin{equation}\label{gaugkinL}
{\cal L}_{gau-kin}=\frac{1}{4}\int d\theta^2 f_{AB}(\phi^I){\cal
W}^A\cdot{\cal W}^B+h.c.
\end{equation}
with superfield strengths ${\cal W}_\alpha=-\frac14\overline{\cal
D}^2\left(e^{-V}{\cal D}_\alpha e^V\right)$, $\alpha$ the spinor
index, and the chiral superfield only entering through the gauge
kinetic holomorphic function $f_{AB}$. We do not consider
Fayet-Iliopoulos terms as they seem to be inconsistent with SUGRA
\cite{Komargodski:2009pc}.
\\
Then, for a generic form for $f_{AB}$, the e.o.m. for the $H^i$
superfield is corrected by
\begin{eqnarray}\label{fullEOM2Extragauge}
\fl\partial_i {\cal L}\supset -\frac14\Phi\left[e^{-G/3}\bar
\Phi\left(G_{i\bar IA}\overline {\cal D}\bar\phi^{\bar
I}+G_{iAB}\overline {\cal D}V^B+ G_{iA}\overline {\cal D}\right)+2
G_{iA}\overline {\cal D}\left( e^{-G/3}\bar
\Phi\right)\right]\overline {\cal D} V^A\cr +\frac14f_{AB,i}{\cal
W}^A\cdot{\cal W}^B\,.
\end{eqnarray}
These are automatically subleading in case the $H$ fields develop
large masses. Indeed, among others, these are related to the SUSY
breaking scale through a D-term breaking.
\\
For factorizable models the SUSY covariant derivatives acting on the
$L$ fields do not appear but (\ref{leadingSol}) is no longer a
solution due to the presence of the SUSY covariant derivatives of
vector superfields. Therefore, the two-derivative description is not
valid either. Notice that all terms inside the brackets in
(\ref{fullEOM2Extragauge}) are null for neutral, i.e., gauge
invariant, $H$ fields as all mixed derivatives of $G$ with the
vector fields vanish.
\\
Actually it is only in the case that the $H$ fields are neutral that
our construction is well stated. Indeed, under a gauge
transformation we can mix the $H$ and $L$ sectors such that the
factorization in $G$ is lost. In physical grounds this is also clear
as the fields that are supposed to be mapped out can be sourced back
by its interaction with the gauge sector, not having any kinematical
constraint forbidding this process.
\\
This last observation warns us about the coupling the $H$ sector can
have with the gauge sector from the sigma model ruled by the gauge
kinetic function. In the e.o.m. this is made explicit in the last
term in (\ref{fullEOM2Extragauge}), which, moreover, makes the
$G_i=0$ a non reliable e.o.m. for the $H$ sector. We have,
therefore, that for neutral fields that appear suppressed in the
gauge kinetic function all terms in (\ref{fullEOM2Extragauge}) are
negligible and one can trust the solutions from (\ref{leadingSol}),
which lead to a two-derivative SUSY effective description.
\\
There might be particular situations where the D-term SUSY breaking
turns out to be suppressed, for instance in the LVS studied in
\cite{mioLVS}, and therefore the back-reaction in the $H$ sector is
mild enough to be subleading. Then, as far as for the scalar
potential is concerned, up to the mass level, no suppression in the
gauge kinetic function is needed for a leading SUSY freezing of the
$H$ sector \cite{mioLVS}. However, this does not imply negligible
contributions to the $H$ superfields solutions coming from other
components of ${\cal W}^{\alpha,A}$, e.g., the field strength and
gauginos in the Wess-Zumino gauge, which would induce, in
particular, non suppressed higher order derivative terms for the
vector fields and higher order fermion bilinear for the gauginos. So
contrary to the case studied by Brizi et al. (see also
\cite{conMarcoII}), where these higher order terms are suppressed by
the mass of the $H$ fields and therefore negligible, an approximate
two-derivative SUSY effective theory is only realized for suppressed
dependencies of the $H$ fields in the gauge kinetic function.
\\
We can allow the $H$ fields to be charged under a hidden gauge
sector ${\cal G}_H$, such that the whole gauge group is given by
${\cal G}={\cal G}_H\otimes {\cal G}_L$ and the vector superfields
are split as $V^a\in{\cal G}_L$ and $V^r\in{\cal G}_H$, labeled by
lower case letters in the beginning and middle of the alphabet
respectively. This avoids easily all possible pathologies we just
mention for a charged $H$ sector.
\\
In this case, however, one should keep in mind the gauge invariance
of the hidden sector, implying that
\begin{equation}
G_r=-iX^i_rG_i\,,
\end{equation}
with $X_r^i$ the Killing vectors, so the set of equations
$\{G_i=0\}$ is no longer linear independent and able to stabilize
all $H$ directions. Indeed, these flat directions are related to
would-be Goldstone fields appearing after gauge symmetry breaking.
We should, therefore, integrate out also the vector superfields that
acquire masses in the process, a situation that can be easily
studied using the superspace approach. With no loss of generality,
in order to be more explicit, we show the Abelian case for which the
e.o.m. reads:\footnote{To obtain this equation we use again
(\ref{DtoF}) but now to write the F-terms as D-terms integrals over
the whole superspace and using the supercovariant derivatives
already present in the superfield strengths.}
\begin{equation}\label{vectorEOM}
G_re^{-G/3}\bar \Phi \Phi+\frac{1}{8}\left[ {\cal D}^{\alpha}\left(f_{rp} \overline {\cal D}^2 {\cal D}_\alpha V^p\right)+\overline{\cal D}^{\dot \alpha}\left(\bar f_{rp} {\cal D}^2 \overline{\cal D}_{\dot \alpha} V^p\right)\right]=0\,,
\end{equation}
where $\alpha$ and $\dot \alpha$ here stand for the spinor index and
we have regarded no kinetic mixing in the gauge sector, i.e.,
$f_{ar}=0$.
\\
Then, requiring the $H$ field dependencies of the gauge kinetic
function for ${\cal G}_L$ to be suppressed, only the SUSY covariant
derivatives on the $H$ and ${\cal G}_H$ sectors appear in the e.o.m
for the $H$ fields. The same should be imposed for the dependency of
the ${\cal G}_H$  gauge kinetic function on the $L$ fields,
otherwise their covariant derivative would appear in the e.o.m. in
(\ref{vectorEOM}). The implementation of the vector superfield
integration corresponding to broken symmetries requires a gauge
fixing, being the unitary gauge the one with clearest physical
interpretation. However, in practice it is useful to work in a gauge
where a chiral superfield, with no vanishing component in the
would-be Goldstone direction, is simply fixed to its VEV. Then, as
long as the SUSY covariant derivatives on the $H$ and $V^r$
superfields are negligible there is a reliable two-derivative SUSY
description after the integration of the fields through the set of
e.o.m.
\begin{equation}\label{chiraplusvector}
G_{\tilde i}=0\,,~~~G_r=0\,,
\end{equation}
where $\tilde i$ runs over the chiral fields not affected by the
gauge fixing, the integration of other fields being encoded in the
longitudinal modes of the massive vector supermultiplets. Then, from
these equations at leading order in $\epsilon$ we have: the lowest
components impose simultaneous F and D-flatness conditions, obtained
by arrangement of the lowest components of the chiral and vector
superfields. The other components are again trivially solved for
vanishing, spinor, vector and auxiliary components, plus their
space-time derivatives. We can be more precise by working in the
Wess-Zumino gauge. In this case the $\tilde i$ index in the set of
equations is not constrained but we can easily spot the effect of
the real superfield components. Again the trivial null components is
solution and the D-flatness condition is obtained, instead, through
the lowest components of the chiral fields alone (see
\ref{EOMComps}).
\\
Notice, that although no restriction on the space-time variation of
the vector and gauginos are obtained, the only possibility of having
vanishing solutions everywhere is because these are also null.
Therefore, the whole supercovariant derivative on the vector
superfield can be neglected as well and the solution to the set of
equations (\ref{chiraplusvector}) are solutions to the e.o.m., which
moreover result in a reliable two-derivative SUSY effective
description.

\section{Gravitational sector and gauge fixing}

In the previous analysis we disregarded the gravitational sector
contribution to the action encoded in the ellipses in
(\ref{SUGRAL}). On the other hand, we have shown that the effective
theory at next to leading order in $\epsilon$ is described by a
theory with superconformal symmetry, namely, the one obtained by the
$G$ function with the $H$ superfields frozen out. Therefore, since
the gravitational terms are univocally dictated by the covariance of
the symmetries these terms are also well described by the truncated
theory.
\\
On the other hand, the dilatation, axial and S transformations of
the superconformal algebra, not being actual symmetries of SUGRA,
should be gauge fixed requiring a canonical normalization in the
gravity sector action, eliminating for example kinetic mixings with
the matter sector. This proceeds by fixing the compensator in terms
of the chiral superfields. However, since the form of $G$ implies
decoupling only between the $H$ and $L$ sector but not with the
compensator, it is not automatically clear that the gauge fixing is
the same in both descriptions or, in other words, that the SUGRA
theory that is obtained upon the gauge fixing coincide at leading
order. Writing the compensator components as
$\Phi=\phi\{1,\chi_\phi,U\}$ the fixing reads \cite{Kugo:1982mr}:
\begin{equation}\label{kugogauge}
\phi\equiv e^{G/6}\,,~~\chi_\phi\equiv \frac13 G_I\chi^I\,,
\end{equation}
where the $G$ function and its derivatives are evaluated in the
lowest components of the superfields and $\chi^I$ are the spinor
components of the chiral multiplets. Since around the solution to
the e.o.m. for the $H$ fields the terms not appearing in the
truncated theory, namely, $G_i\chi^i$, are of order $\epsilon$ and
the functions $G$ and $G_\alpha$ coincide in both theories at next
to leading order, the gauge fixing is the same modulo subleading
terms.
\\
One of the main targets of the present letter is to clarify the
integration of the fields at the superfield level, however, the
gauge fixing in (\ref{kugogauge}) cannot be promoted to the
superspace as the compensator is a chiral field and therefore cannot
depend on the fields in the antiholomorphic sector contained in $G$.
A variation to the fixing which can be performed directly in the
superspace is the one proposed by Cheung et al. in
\cite{Cheung:2011jp} that in our K\"ahler gauge reads:
\begin{equation}
\Phi\equiv e^{Z/3}(1+\theta^2 U)\,,
\end{equation}
with $Z$ a chiral superfield given by
\begin{equation}
Z=\langle G\rangle+\langle G_I\rangle\phi^I\,,
\end{equation}
where the $\langle\rangle$ means the VEV. Again since the VEV's in
both descriptions coincide at leading order and the terms not
appearing in the truncated description are suppressed, the $Z$
superfields, and therefore the full and truncated theories, match at
leading order.

\section{Discussion}

In this letter we have studied the possibility of having a SUSY
two-derivative description for effective theories resulting from the
integration of light fields in ${\cal N}=1$ SUGRA. The consistency
of a derivative expansion with SUSY transformations requires a
parallel expansion in spinor bilinears and auxiliary terms, that at
the superfield level is seen as an expansion in the supercovariant
derivatives and a reliable two-derivative effective description is
the one where these can be neglected.
\\
A first point drawn in the paper is the possibility, in a generic
fields theory, of integrating out light fields provided there is a
decoupling between the different modes. This decoupling serves as
dynamical constraint that keeps the modes that have been mapped out
indeed out of the theory although not kinematical constraint is
available. This leads to what we call effective descriptions where
we further allow a small coupling between the sectors. A first
instance of this kind of situation is worked out in a rigid SUSY
example. Here we explore also the condition for the two-derivative
effective description to be consistent with SUSY transformations.
Interestingly enough the moral learned here can be extended to the
SUGRA situation explored later on.
\\
Whenever we speak about an effective description we have in mind a
region in the field configuration space around particular solutions
of the e.o.m. for the fields that have been integrated out. We find
that the integration of superfields leads to a reliable
two-derivative SUSY effective description if such solutions preserve
SUSY, approximately, albeit the remaining fields stand at points
where SUSY is spontaneously broken. One possibility is that the SUSY
preserving sector is heavy enough to present a hierarchy with the
SUSY breaking scale such that the back-reaction from the breaking is
suppressed \cite{conMarcoI,Brizi:2009nn,conMarcoII}. For K\"ahler
potentials with no singular behavior such a hierarchy is realized if
in particular the gravitational effects, e.g., the contribution to
the masses, are suppressed, and therefore the leading superfield
e.o.m. coincides with the one obtained in rigid SUSY.
\\
On the other hand, no hierarchy is necessary if in the Lagrangian
the two sectors are decoupled, in which case the best scenario in
SUGRA would be sequestered sectors. Still, one can allow further
interactions, beside the gravitational ones, and achieve some SUSY
decoupling if the theory is described by a K\"ahler invariant $G$
function of the form (\ref{factG}). Although this was previously
realized at the level of the scalar Lagrangian
\cite{Achucarro:2008sy,conMarcoI,mioLVS}, our analysis shows that
the situation can be understood in a fully SUSY framework by working
directly in the superspace, an approach that also allows the study
of more involved situations not regarded before. We find that the
decoupling leads to subleading contributions from the SUSY covariant
derivatives on the $L$ sector, despite the fact these can be large,
and then field configurations solving (\ref{leadingSol}) coincide at
leading order, in a $\epsilon$ expansion, with solutions of the
exact e.o.m., implying negligible supercovariant derivatives from
the $H$ sector and, therefore, a reliable two-derivative SUSY
description. Nicely enough this can be summarized as the condition
that the superfield equation obtained as analytical continuation of
the F-flatness condition to the superspace be a reliable e.o.m. for
the $H$ sector, independently of the $L$ sector dynamics.
\\
This superfield equation, although can be seen as a natural and
naive guess, was already criticized as e.o.m. for chiral fields.
Indeed, the equation is a real composite superfield equation, so it
overdetermines a chiral solution having more equations than
unknowns. However, for the case heavy $H$ fields, compared with the
SUSY breaking scale, this extra terms turn out to be negligible and
the equation takes the chiral form found in \cite{Brizi:2009nn}
given by (\ref{heavycase}). On the other hand if the theory has a
factorizable nature the equation continues to be a real and in
particular the gravitational effects are important. However, it is
trivially solved at leading order by null higher components of the
chiral superfield, and therefore it is consistent with a chiral
solution.
\\
It is important to notice that, contrary to the sequestered case,
for the factorizable models it is only around the SUSY
configurations for the $H$ sector that there is freedom on
integrating and decoupling light fields. Indeed, in case the $H$
sector leads the breaking of SUSY the decoupling with the $L$ sector
is lost, as can be seen in (\ref{leadEOMwithep}), and the e.o.m. for
$H$ starts to be $L$ dependent. In this case, although nice
decoupling features are preserved
\cite{Achucarro:facto,Achucarro:2010jv}, some constraints on the
mass of the integrated fields appear for a decoupling to apply
\cite{Hardeman,Postma}. Interestingly enough, under these
circumstances for the effects of integrating out fields requiring a
reliable two-derivative SUSY description we find that the
factorizable models are as safe as the sequestered ones.
\\
The fact that ours is not a low-energy description alerts about the
fact that even in case the fields to be integrated out are neutral
these can be sourced back by the vector fields from the coupling in
the sigma model ruled by the gauge kinetic function. Thus, even if
the D-term SUSY breaking is mild other terms in the covariant
derivative of the gauge vector are not suppressed and therefore no
reliable SUSY two-derivative description is available. One should,
then, require a suppressed dependency of the gauge kinetic function
on the $H$ fields and, in this case, all the covariant derivative
contributions to the e.o.m. are negligible at leading order such
that the solutions are determined by (\ref{leadingSol}).
\\
We can allow charged $H$ fields but only under some hidden gauge
group in which case the $L$ fields should appear in a suppressed way
in the corresponding gauge kinetic function. Possible flat
directions resulting from symmetry breaking are handled by
integrating out the vector supermultiplets, after a gauge fixing for
the broken directions.
\\
A nice fact of working directly in the superspace is that we are
able to spot further considerations in the full set of e.o.m. that
are cumbersome to find in working with the component Lagrangian. So,
for example, we find that the condition for slow varying
configurations in the $H$ sector for a reliable two-derivative SUSY
description is actually contained in the e.o.m. and there is no need
of imposing it from outside, as was argued in \cite{mioLVS}. That
the e.o.m. induces vanishing space-time derivatives can be
understood from the fact that it is the only way to preserve SUSY,
since otherwise these would induce non-vanishing values of the
spinor and auxiliary fields after SUSY transformations.
\\
Then, in general, we can say that the theory obtained by integrating
out fields is reliably described by a two-derivative SUSY theory if
the analytic extension of the F-flatness and D-flatness conditions
to the superspace are reliable superfield e.o.m. for the process of
integration. In other words, SUSY preserving solutions is a
sufficient condition for such a description to proceed. Clearly, in
concluding this we have in mind that all operators that appear
suppressed before the truncation have a counterpart in the truncated
Lagrangian, otherwise will be misleading to neglect them in some
contexts, like is the case of baryon number violating operators in
effective descriptions of grand unified theories and the study of
proton decay.
\\
In order to be more precise about the validity of these results one
should also look for possible energy and momentum transfer from the
background, dictated by the solutions for $H$, to the the $L$
sector. These, in our case indeed happen as the space-time variation
in $H$ sector is not exactly null. However, being of order
$\epsilon$ we can safely scan regions in the space-time that are
within a radius of order $\epsilon^{-1}$, out of which we cannot
longer consider the $H$ sector as homogeneous and in particular SUSY
preserving.
\\
Then, although our study is a step forward in the understanding one
of the many simplification behind explicit constructions in
supersymmetric theories it would be important to explore more
carefully the nature of higher order operators. Another immediate
question is to what extend our conclusions apply in more complicated
setups, like ones that mix mild hierarchies and rather small
couplings simultaneously, and/or with further sectors in the game.
Those questions we hope to address in future works.

\section*{Acknowledgements}
I benefited from conversations with Leonardo Brizi, Francesco
D'Eramo, Jorge Zanelli and Marco Serone. I thank specially Claudio
Scrucca and Marco Serone for useful comments on a preliminary
version of the paper. Important and useful comments from an
anonymous referee are also acknowledged.
I would like to thank The Abdus Salam International Centre for Theoretical Physics for the hospitality during the completion of this work.\\

\appendix

\section{Components for the equation $G_{H,i}=0$.}\label{EOMComps}
Let us start with the case of global SUSY for which the leading
e.o.m. is $W_{H,i}=0$. To make explicit the components we write it
as
\begin{equation}
W_{H,i}(H)=W_{H,i}(h)+W_{H,ij}(h)\Delta^j+\frac12
W_{H,ijk}(h)\Delta^j \Delta^k\,,
\end{equation}
where $\Delta^i=H^i-h^i=\sqrt{2}\theta\psi^i-\theta \theta F^i$,
$h^i$, $\psi^i$ and $F^i$ the lowest, spinor and auxiliary
components of the superfield $Hî$, and the argument in the fields is
$y^\mu=x^\mu+i\theta \sigma^\mu \bar \theta$. Expanding to make
explicit the $\bar \theta$ dependencies so far implicit in $y^\mu$,
we find the components:

\begin{eqnarray}
\theta^0~~~~~:W_{H,i}=0\,,\\
\theta~~~~~~:  W_{H,ij}\psi^j=0\,,\\
\theta \sigma^\mu \bar\theta~~\,\,: W_{ij}\partial_\mu h^j=0\,,\\
\theta^2~~~~~~\,:-  W_{H,ij}F^j-\frac12 W_{H,ijk}(h)\psi^i \psi ^j=0\,,\\
\theta \sigma^\mu \bar\theta \theta_\alpha: \partial_\mu
(W_{ij}\psi^\alpha)=0\,,\\
\theta^2\bar\theta^2~~~~: W_{ijk}\partial^\mu h^j \partial_\mu
h^k=0\,.
\end{eqnarray}
The first equation implies that the lowest components arrange so to
satisfy the F-flatness condition. The others are then solved with
vanishing spinor and auxiliary components. Moreover, vanishing
space-time derivatives are required from the non-holomorphic
components.
\\
For $G_{H,i}$ we can do something similar but now expanding $\Delta$
making explicit the $\theta$ dependencies. Then,
\begin{eqnarray}
\theta^0~:G_{H,i}=0\,,\\
\theta~~:  G_{H,ij}\psi^j=0\,,\\
\bar\theta~~: G_{H,i\bar j}\overline{\psi}^{\bar j}=0\,,\\
\theta^2~: -  G_{H,ij}F^j-\frac12 G_{H,ijk}\psi^j \psi
^k=0\,,\\
\bar\theta^2~: -  G_{H,i\bar j}\overline F^{\bar j}-\frac12
G_{H,i\bar
j\bar k}\overline \psi^{\bar j} \overline \psi^{\bar k}=0\,,\\
\theta \sigma^\mu \bar\theta:~i G_{H,ij}\partial_\mu h^j-iG_{Hi\bar
j}\partial_\mu \bar h^{\bar j}+G_{H,ij\bar k}\psi^j\sigma_\mu
\overline \psi^{\bar k}=0\,,
\\
\bar\theta \theta^2~: -i\left[G_{H,ij}\partial_\mu \psi^j-\frac12
G_{H,ijk}\left(\partial_\mu h^j\psi^k+\partial_\mu
h^k\psi^i\right)-G_{H,ij\bar k} \partial_\mu \bar h^{\bar
k}\psi^j\right]\sigma^\mu\cr~~~~~~~~-2G_{H,ij\bar
k}F^i\overline\psi^{\bar k}- G_{H,ijk\bar
l}\psi^j\psi^{k}\overline \psi^{\bar l}=0\,,\\
\theta \bar\theta^2~: i \sigma^\mu\left[G_{H,i\bar j}\partial_\mu
\overline\psi^{\bar j}-\frac12G_{H,i\bar j\bar k}\left(\partial_\mu
\bar h^{\bar j}\overline\psi^{\bar k}+\partial_\mu\bar h^{\bar
k}\overline \psi^{\bar i}\right)-G_{H,ij\bar k}
\partial_\mu h^{j}\overline\psi^{\bar k}\right]\cr~~~~~~~~-2G_{H,ij\bar
k}\overline F^{\bar k}\psi^{j}-G_{H,ij\bar k\bar l}\psi^j\overline
\psi^{\bar k}\overline \psi^{\bar l}=0\,,\\
\theta^2\bar \theta^2~:G_{H,ij\bar j}\left( F^j\overline F^{\bar
j}+\partial_\mu h^j\partial^\mu\bar h^{\bar
j}-\frac{i}{2}\psi^j\sigma^\mu
\partial_\mu \overline\psi ^{\bar
j}+\frac{i}{2}\partial_\mu\psi^j\sigma^\mu  \overline\psi ^{\bar
j}\right)\cr ~~~~~~~ +\frac{i}{4}G_{H,ijk\bar
k}\left(\psi^j\sigma^\mu\overline\psi^{\bar k}\partial _\mu
h^k+\psi^k\sigma^\mu\overline\psi^{\bar k}\partial _\mu
h^j-2ih^j\psi^k\overline F^{\bar k}\right)\cr ~~~~~~~
+\frac{i}{4}G_{H,i\bar j\bar k k}\left(
\psi^k\sigma^\mu\overline\psi^{\bar j}\partial _\mu \bar h^{\bar
k}+\psi^k\sigma^\mu\overline\psi^{\bar k}\partial _\mu \bar h^{\bar
j}+2i \bar h^{\bar j}\overline\psi^{\bar k}F^{k}\right)\cr
~~~~~~~+\frac14G_{H,ijk\bar j\bar k}\psi^j\psi^k\overline\psi^{\bar
j}\overline\psi^{\bar k}-\frac14\partial_\mu \partial^\mu G_{H,i}
=0\,,
\end{eqnarray}
where the arguments for $G_{H,i}$ and its field derivatives is the
lowest components $h$ and $\bar h$. The first equation indicates
that the lowest components are arranged such to solve the F-flatness
condition. The higher order components imply vanishing spinor and
auxiliary components. Vanishing space-time derivatives are also
necessary for the solution.
\\
The presence of vector superfields change the previous components by
adding the following terms, in the Wess-Zumino gauge:
\begin{eqnarray}
\theta \sigma^\mu \bar\theta:~G_{H,is}v_\mu^s\,,\\
\bar \theta \theta^2: iG_{H,is}\bar\lambda^s_{\dot
\alpha}-\frac{1}{\sqrt{2}}G_{H,ijs}v^s_\mu (\psi^j \sigma^\mu)_{\dot
\alpha}\,,\\
\theta \bar\theta^2: -iG_{H,is}\lambda^s_{
\alpha}-\frac{1}{\sqrt{2}}G_{H,ijs}v^s_\mu (\sigma^\mu\overline
\psi^j )_{\alpha}\,,\\
\theta^2\bar\theta^2:\frac12 G_{H,is}D^s+\frac14
G_{H,ist}v_\mu^sv^{\mu,t}\cr~~~~~~~+\frac{i}{2}G_{H,ijs}(\partial_\mu
h^jv_\mu^s-\sqrt{2}\psi^j\bar \lambda^s)-\frac{i}{2}G_{H,i\bar
js}(\partial_\mu \bar h^{\bar j}v_\mu^s-\sqrt{2}\lambda^s \bar\psi^j
)\,.
\end{eqnarray}
The superfield e.o.m. $G_r=0$ has analogous leading terms, obtained
just by changing in the previous ones the index $i$ by $r$.

\section*{References}
\bibliographystyle{iopart-num}

\providecommand{\newblock}{}

\end{document}